# Understanding Life with Molecular Dynamics and Thermodynamics:
*Comment on **Nature** 451, 240-243 (2008)*

Bob Eisenberg

beisenbe@rush.edu
+1-312 942 6467

Dept of Molecular Biophysics and Physiology
Rush University Medical Center
Chicago IL 60612
USA

February 15, 2008

# Comment on Nature News Article on Molecular Dynamics

Dear Sir or Madam,

The news in the recent article ("Chemistry Power Play": Nature 451, 240-243 (2008)) describing the wonderful advances in molecular dynamics MD is most welcome to those of us who try to understand how the atomic structure of ionic channels control biological function like the currents that flow through channels and control signaling in the nervous system and contraction in the heart. Schulten's group[1] (along with others[2]) has shown how these currents can be computed accurately and efficiently from MD simulations even though current is a macroscopic variable that flows on a millisecond time scale in the heart and nerves. (The time scale of MD is about 0.1 femtoseconds.)

Biological functions of ion channels, enzymes, and other proteins depend sensitively on macroscopic (thermodynamic) variables like current, concentration and membrane potential — as any physiology or biochemistry textbook illustrates — and these are difficult to understand or compute with reasonable accuracy using the MD packages mentioned in the article, as they exist now. New methods are needed, in my view. Many of these difficulties arise because the MD packages mentioned do not describe the thermodynamic properties of ions in solutions very well (e.g., free energy per mole, called the activity).

Simulations of channels that include detailed description of the ions, but not of the protein, are surprisingly successful. For example, a simple model[3] with only two parameters — **whose values are set once and never adjusted** — can account for the selectivity of the DEKA (Asp-Glu-Lys-Ala) sodium channel in many solutions of different ions with compositions varying over many orders of magnitude, even though the model uses only information about the primary structure, namely, the charge and excluded volume of the important amino acid side chains. The same model — **without adjusting any parameters** — accounts for the properties of the mutant channel DEEA (Asp-Glu-Glu-Ala), in a wide range of conditions, even though DEEA is a calcium channel with very different properties from a Na channel.

One of the important challenges facing high resolution MD calculations is to reproduce the success of such embarrassingly simple and successful models. This challenge must be faced, in my view, because many biological functions occur in the macroscopic world and involve macroscopic variables. We are all eager to see how molecular dynamics will accurately estimate the thermodynamic variables that define and control so many macroscopic biological functions.

Bob Eisenberg
Department of Molecular Biophysics
Rush University Medical Center
Chicago IL 60612 USA
beisenbe@rush.edu
+1-312 942 6467

February 15, 2008